# Clean BN encapsulated 2D FETs with lithography compatible contacts

Binxi Liang[1,2,†], Anjian Wang[1,2,†], Jian Zhou[1,2], Shihao Ju[1,2], Jian Chen[1,2,*], Kenji Watanabe[3], Takashi Taniguchi[3], Yi Shi[1,2], Songlin Li[1,2,*]

*[1] National Laboratory of Solid-State Microstructures and Collaborative Innovation Center of Advanced Microstructures, Nanjing University, Nanjing, Jiangsu 210023, China*

*[2] School of Electronic Science and Engineering, Nanjing University, Nanjing, Jiangsu 210023, China*

*[3] National Institute for Materials Science, Tsukuba, Ibaraki 305-0044, Japan*

[†]These authors contributed equally to this work.

*Correspondence should be addressed to J.C. (chenj63@nju.edu.cn) or S.L. (email: sli@nju.edu.cn).

**ABSTRACT:** Device passivation through ultraclean hexagonal BN encapsulation is proven one of the most effective ways for constructing high-quality devices with atomically thin semiconductors that preserves the ultraclean interface quality and intrinsic charge transport behavior. However, it remains challenging to integrate lithography compatible contact electrodes with flexible distributions and patterns. Here, we report the feasibility in straightforwardly integrating lithography defined contacts into BN encapsulated 2D FETs, giving rise to overall device quality comparable to the state-of-the-art results from the painstaking pure dry transfer processing. Electronic characterization on FETs consisting of $WSe_2$ and $MoS_2$ channels reveals an extremely low scanning hysteresis of ~2 mV on average, a low density of interfacial charged impurity of ~$10^{11}$ $cm^{-2}$, and generally high charge mobilities over 1000 $cm^2V^{-1}s^{-1}$ at low temperatures. The overall high device qualities verify the viability in directly integrating lithography defined contacts into BN encapsulated devices to exploit their intrinsic charge transport properties for advanced electronics.

## INTRODUCTION

With atomic-scale thickness and high charge mobility ($\mu$), atomically thin two-dimensional (2D) semiconductors, such as transition-metal chalcogenides (TMCs), are considered as promising candidates for electronics at technological nodes beyond silicon.[1–3] However, charge carriers in 2D semiconductors are highly sensitive to extrinsic scattering sources located at interfaces such as Coulomb impurities due to their extremely reduced thickness.[4] Hence, effective passivation strategies that can preserve ultraclean interface quality and intrinsic electronic performance are extensively investigated. Among them, device passivation through BN encapsulation remains one of the most effective ways for constructing ultraclean devices consisting of 2D TMC channels, which enable probing their intrinsic charge transport behavior and evaluating the potential for electronics at advanced technological nodes.

In case of TMC channels, contact integration in BN passivated devices represents an intractable step, because the semiconducting nature prevents using the edge contacts directly, as performed for graphene.[5,6] Thus, many efforts have been devoted to integrate the contact electrodes into the BN encapsulated/passivated channels.[7–13] In an important advance, Hone et al. first reported the multi-terminal integration strategy by sequential transfer of individual graphene electrodes onto TMC channels before final BN encapsulation.[7,8] In a simplified version that requires transferring contact flakes once, Kretinin et al. devised a smart means by wiring out multiple Hall electrodes through detoured wiring paths after post-patterning the contact and channel flakes.[9] Alternatively, Wang et al. developed a more direct way based on finely etching the local BN areas to expose the channel areas underneath.[10,11] Recently, Hone et al further updated their integration strategy by directly transferring the top BN encapsulator with predefined perforating contacts onto the TMC channels.[12,13] All the strategies above can result in satisfied passivation effect for TMC channels including ultraclean channel interfaces and high-quality contacts, but most lack processing simplicity, convenience or high yields because they generally evolve multiple lithography and etching processing steps.

In this work, we report a straightforward contact integration strategy into BN encapsulated TMC devices, by directly transferring BN supported channels onto lithography prepatterned contacts, and the overall high device qualities. We show that the cleanness of the bottom BN surface can be well maintained through appropriate thermal annealing after the straightforward metallization. Since the contacts are defined by lithograph, there are basically no strict limitations on their shapes and dimensions. Well encapsulated field-effect transistors (FETs) with two typical TMC channels including unipolar $MoS_2$ and ambipolar $WSe_2$ are fabricated with this strategy to verify the realistically high device qualities. An extremely low average scanning hysteresis of ~2 mV, a low density of interfacial charged impurity of ~$10^{11}$ $cm^{-2}$, and generally high low-temperature charge mobilities over 1000 $cm^2V^{-1}s^{-1}$ are observed, making them among the typical range of BN encapsulated counterparts. Besides, a low contact resistance of 20 k$\Omega\cdot\mu$m and a low effective thermal contact barrier of 34 meV are extracted for the $MoS_2$ devices, indicating the excellent contact quality between channels and electrodes defined by van der Waals stacking. This straightforward contact integration protocol provides optimum balance between device performance and processing simplicity.

## EXPERIMENTAL

Figure 1a shows the schematic fabrication steps for integrating the patternable contacts into BN encapsulated FETs consisting of semiconducting 2D TMC channels. First, the contact electrodes (3 nm Ni, 12 nm Au) with designed patterns are defined onto bottom BN encapsulator by standard electron beam lithography (EBL), followed by appropriate thermal annealing (250 °C for 30 min in mixed $Ar/H_2$ atmosphere) to remove the possible resist residues from lithography, as shown in Fig. 1b. Then, the TMC channel is picked up with van der Waals method by top BN encapsulator (Fig. 1c). Afterwards, the top stacks are transferred onto the metallized bottom BN encapsulator (Fig.1a) to fully encapsulate the TMC channel with multiple contacts wiring outwards (Fig. 1d). Finally, the top gate is defined to accomplish the fabrication of the encapsulated FET (Fig.1e). In this study, two typical TMC channels including unipolar $MoS_2$ and ambipolar $WSe_2$ are fabricated to verify the realistically high device qualities of this straightforward contact integration strategy.

Figure 1f shows the surface morphology for a typical corner of the fabricated $WSe_2$ FET, as revealed by atomic force microscopy (AFM), to analyze the characteristic thickness information of the channel and dielectric flakes employed, where the $WSe_2$ channel, top and bottom BN encapsulators are ~1.5, ~26, and ~12 nm, respectively, as identified by the AFM profile lines (Fig. 1g). The channel thickness indicated it is bilayer (2L). We also check the effect of passivation of this integration strategy by comparing the Raman and photoluminescence (PL) spectra for the channel before encapsulation and after storage in ambient for one month (Fig. 1 h and i). In both cases, the peak positions are nearly identical after long-time storage, proving the validity of device passivation. Furthermore, the Raman peak becomes narrower after encapsulation (Fig. 1h), indicating a cleaner interfacial condition.[14] The synergic improvement in interfacial condition and ambient stability after encapsulation lays the foundation in achieving the intrinsic performance of the 2D TMC channels.

## RESULTS AND DISCUSSION

To gain information on the realistic device quality from the straightforward contact integration strategy, we systematically checked or analyzed multiple device parameters including scanning hysteresis, residual trap density, contact quality and cryogenic $\mu$. Figure 2a shows the typical transfer characteristics for an encapsulated FET with 6-layer (6L) $MoS_2$ as channel at 300 and 7.5 K. During measurement, the top gate was swept forwards and backwards between -3 to 8 V with a 32-nm BN as dielectric, to check the scanning hysteresis, and the Si bottom gate is grounded to avoid additional capacitive coupling. The FET exhibits typical unipolar n-type conduction behavior at both temperatures ($T$s). At 300 K, a high on-off ratio more than $10^7$ and a low subthreshold swing (SS) of 83 mV/dec are observed while at 7.5 K they are further improved to ~$10^9$ and 42 mV/dec, respectively. Due to the high interfacial quality of BN dielectrics, the extrinsic doping effect from dielectrics to channels is minimized and, thus, the 6L $MoS_2$ channel is nearly neutral with threshold voltage ($V_t$) around -1V at 300 K. For practical applications, this small $V_t$ deviation is expected to be readily tuned through various engineering strategies, such as modulation of the work function of metal gates.[15]

Remarkably, the curves show negligible scanning hystereses in the full-range plot and they become discernible only in the enlarged plots over a 10-mV range, as shown in the insets of Fig. 2a. Their values are almost independent with $T$ from 300 to 7.5 K with an average value of 2 mV, as summarized in the inset of Fig. 2b. It is commonly believed that the scanning hystereses originate from the capture and release of the trap states located at the dielectric/channel interfaces.[16,17] The density of such trap states, strongly associated with device cleanness, can be estimated with the formula $\Delta Q = C_{tg} \cdot \Delta V_H / e$, where $C_{tg}$ is the dielectric capacitance per unit area, $\Delta V_H$ is the scanning hysteresis and $e$ is the elementary charge. Figure 2b compares the $\Delta Q$ value of our fully encapsulated $MoS_2$ FET with those from single-supported, unencapsulated FETs, including BN, $SiO_2$ and $Al_2O_3$ dielectrics.[16,17] As compared with the values in unencapsulated FETs, the density of trap states in the fully encapsulated FETs are generally 2–3 orders lower in magnitude, confirming the preservation of the ultraclean nature of the van der Waals interfaces of layered BN and $MoS_2$.

To gain insight into the contact quality, we further extracted the effective Schottky barrier height ($\phi_B$) of $Au/MoS_2$ contacts through the thermionic emission (TE) theory[18-20] and the dependence of contact resistance ($R_c$) on $T$. It is widely accepted that below flat-band voltage ($V_{FB}$), the charge injection into 2D materials is mainly dominated by the TE mechanism and the channel current $I_{ds}$ can by expressed as[18,21,22]

$$I_{ds} = A_{2D}^* T^{3/2} \exp\left(-\frac{q\phi_B}{k_B T}\right), \qquad (1)$$

where $A_{2D}^*$ is the equivalent Richardson constant, $T$ is the temperature, $k_B$ is the Boltzmann constant and $q$ is the elementary charge. Figure 2c shows the plot of $\ln\left(I_{ds}/T^{3/2}\right)$ versus $1000/T$ at different gate voltages and the slopes are normally extracted as the effective $\phi_B$ under each gating conditions. Figure 2d summarizes the extracted $\phi_B$ at different $V_{TG}$, where the point deviating the linear trend is normally adopted as $V_{FB}$. As such, the barrier height at the flat-band condition ($V_{FB}$=-0.6 V) is about 34 meV, which is close to the RT thermal energy. The inset in Figure 2d demonstrates contact resistance ($R_C$), extracted by 4-probe method, as a function of $T$ from 300 to 7.5 K. The values are comparable with the counterparts with contacts from conventional techniques[23], indicating the excellent contact quality made from the van der Waals stacking method. When $V_{TG}$>-0.6 V, negative $\phi_B$ values are extracted, suggesting the transition of injection mechanism from TE to thermally assisted tunneling and the failure of the TE mechanism above $V_{FB}$.[20,24]

Next, we investigate the cryogenic electronic transport properties of FETs comprising two typical TMC channels including the unipolar 6L $MoS_2$ and another ambipolar 2L $WSe_2$. Figure 3a,b shows the 2-probe transfer characteristics of the unipolar $MoS_2$ FET at $V_{ds}$=1 V in logarithmic and linear plots, respectively. Likewise, the transfer characteristics of the ambipolar WSe$_2$ FET are given in Figure 3d,e. In both cases, near-zero hystereses are observed at all measured $T$s, confirming again the preservation of interface cleanness after the encapsulation. In addition, on-off current ratio increases as $T$ decreases. For instance, it increases from $10^7$ to $\sim 10^8$ in the electron regime of the WSe$_2$ FET when $T$ is reduced from 300 to 8 K. The negative correlation between switching ratio and $T$ is a consequence of high overall device quality including interface cleanness and Ohmic contacts. Even in case of the bipolar WSe$_2$ FET, this trend remains valid (Fig. 3e), in spite of smaller $T$ effect than MoS$_2$ due to enhanced $\phi_B$ for both the hole and electron injection.

The dependence of $\mu$ on $T$ is also carefully analyzed to evaluate the passivation effect of this strategy. To exclude the contact resistance and extract the intrinsic $\mu$ values, the 4-probe method is adopted. Figure 3c shows the dependence of $\mu$ on $T$ for the n-type MoS$_2$ FET (aspect ratio $\sim$ 0.64). $\mu$ reaches 110 and 4270 cm$^2$V$^{-1}$s$^{-1}$ at 300 and 7.5 K, respectively, which are among the typical values of the BN encapsulated devices.[7,8,10] For the ambipolar WSe$_2$ FET, there are two sets of $\mu$ values for hole and electron. The corresponding values are given in Fig. 3f. On the side of electron, $\mu$ reaches 71 and 1078 cm$^2$V$^{-1}$s$^{-1}$ at 300 and 7.5 K, respectively. In contrast, they are 232 and 1521 cm$^2$V$^{-1}$s$^{-1}$ on the side of hole. Thus, the $T$ dependence is stronger for electron injection. The observation of generally high $\mu$ values over 1000 cm$^2$V$^{-1}$s$^{-1}$ at low $T$ for the both TMC materials indicates the preservation of high interface cleanness.

For the atomically thin TMC channels, the charge scattering mechanisms can be mainly attributed to lattice phonons, remote interface phonons (RIPs), charged impurities (CIs) at interfaces, and lattice vacancies,[2,4,7,25] where the former two factors are highly $T$-dependent and become activated primarily at high $T$ regime, because of the thermal nature of various phonons, while the latter two are less $T$-dependent. Thus, it is informative to evaluate the qualities for device passivation and material crystallinity by identifying the leading charge scattering sources at different $T$ regimes.

The trends of $\mu$ versus $T$ shown in Fig. 3 c,f exhibit typical band-like transport behavior, as expected for well passivated clean samples.[4,16,26] Below 30 K, $\mu$ is weakly dependent on $T$, which is in line with the theoretical predictions on interfacial CIs as leading scattering sources. The interfacial cleanness can be directly estimated with the density of CIs ($n_{CI}$) through theoretical calculations. By fitting the $\mu$–$T$ data,[4] we derived upper limits of $10^{11}$ and $1.1 \times 10^{11}$ cm$^{-2}$ for $n_{CI}$ in the MoS$_2$ and WS$_2$ devices. Such levels are one order lower in magnitude than the pristine Si-SiO$_2$ interface.[27,28]

At the high-$T$ regime from 300 to 50 K where the lattice and remote interface phonons are dominated, resulting in pronounced $\mu$–$T$ dependence. Empirically, a power law ($\mu \propto T^{-\gamma}$) is employed to describe this $\mu$–$T$ dependence and the value of $\gamma$ has been used to evaluate the weights of extrinsic scattering mechanisms such as CI, RIP and lattice vacancies. The higher the weights of extrinsic scattering mechanisms, the lower the $\gamma$ value is. Theoretical studies on the case of pure lattice phonon scattering predicted $\gamma \sim 1.7$ for 1L MoS$_2$.[29] In experiment, values of $\sim 1.2$[8] and 1.9[7] are reported for graphene contacted, BN encapsulated 1L MoS$_2$, and a value of $\sim 2.6$ was reported for bulk MoS$_2$.[30] For our 6L MoS$_2$ device, the $\gamma$ value is around 1.7, which is generally higher than those of SiO$_2$ supported samples and within the range of high-quality BN encapsulated devices (Fig. 4a). The $\gamma$ values for the electron and hole sides in WSe$_2$ FET are $\sim 1.5$ and $\sim 0.8$ respectively, which fall in the trend of fully encapsulated counterparts (Fig. 4b).[11,12,31] Because of the higher $\gamma$ value, the electron mobility surpasses the hole mobility at cryogenic temperature, even though it is lower at RT. To sum up, the interfaces in our devices show considerably low CI concentration, further proving the straightforward integration of patterned contacts results in less contaminated channel interfaces.

Finally, we compare the $\gamma$ value and cryogenic $\mu$ of our FETs with those in literature to further elucidate the passivation effect of the straightforward integration strategy. We note that $\mu$ in TMC are generally thickness dependent.[4] Thus, the number of layers (NLs) is an important parameter to be clarified for a fair comparison. Figure 4a,c summarizes the $\gamma$ values for MoS$_2$ and WSe$_2$, respectively. In both cases, the devices made by our strategy exhibit comparable, albeit not the highest, values to reported BN encapsulated devices.[7,8,11,12,18,26,32] Meanwhile, as shown in Fig. 4b,d, at low $T$ our devices outperformance those defined by all other passivation methods and exhibit comparable $\mu$ values with those reported in fully BN encapsulated counterparts by pure dry transfer.[7,8,10,26,31–38] Hence, our simplified fabrication process can lead to high charge mobility comparable with the state-of-the-art pure dry transfer processing without much compromise on device performance.

## CONCULSIONS

We proposed a straightforward strategy for integrating lithography compatible contacts into BN-encapsulated TMC devices and demonstrated that the cleanness of the channel interfaces can be preserved to a high degree through appropriate thermal treatment which gives rise to comparably high device performance with other methods relying on dry but complicated transfer processing. Electronic analyses on FETs compromising $WSe_2$ and $MoS_2$ channels uncover high interface cleanness and contact quality, which result in generally high overall electronic performance. This straightforward contact integration protocol represents optimum balance between device performance and processing simplicity.

## Supporting Information

Device current under $V_{FB}$ for extracting injection barrier; extraction condition for contact resistance; injection limit behavior under bottom gated mode

## METHODS

**Lithography and metallization.** Bilayer resists comprising poly methyl methacrylate (PMMA, A4) and methyl methacrylate (MMA, 8.5) was adopted in EBL, to facilitate the following procedure of lift-off, for patterning the contact electrodes on the bottom BN flakes. A cold-field-emission electron source consisting of $LaB_6$ filaments was equipped in the EBL system, with the lithography resolution down to 100 nm. A dose of 230 $\mu C/cm^2$ was employed to slightly overexpose the resist bilayer to minimize the possible resist residues after development in methyl isobutyl ketone solution. Then, the resist patterns of contact electrodes were rinsed in isopropyl alcohol and were blow dried by nitrogen flow. Afterwards, thin Ni/Au (3 nm/12 nm) contact electrodes were defined using a combination of thermal evaporation deposition and standard lift-off. Finally, the bottom BN flakes with the lithography-defined electrodes were thermally annealed at 250 ℃ for 30 min in $Ar/H_2$ atmosphere to further remove the possible resist residues.

**Van der Waals stacking.** To fabricate the BN encapsulated FETs, individual flakes of BN and TMC were initially mechanically exfoliated onto $SiO_2$/Si substrates. Then, a BN flake was selected as top encapsulator and picked up with a thick polydimethylsiloxane (PDMS) slab coated with polypropylene carbonate (PPC) polymer at 45 ℃ in nitrogen glove box. The top BN encapsulator on the PMMA/PPC slab was then used to pick up a few-layer TMC sheet and transfer it to the bottom BN encapsulator with predefined contact electrodes. The PPC on the top side of the BN/TMC/contacts/BN stack was melted at 120 ℃ to release the stack from the stacking slab and was cleaned with acetone. Next, the BN/TMC/contacts/BN stack was annealed at 150 ℃ for 30 min in $Ar/H_2$ atmosphere to remove the bubbles between the layers. Finally, a top gate and outward wiring pads (3nm Ni/ 45 nm Au) were defined together by standard EBL and metallization.

## ACKNOWLEDGMENTS

S. L. acknowledges the supports from the National Key R&D Program of China (2021YFA1202903), the National Natural Science Foundation of China (61974060, 61674080 and 61521001), the Innovation and Entrepreneurship Program of Jiangsu province and the Micro Fabrication and Integration Technology Center in Nanjing University. K. W. and T. T. acknowledges the supports from the Elemental Strategy Initiative conducted by the MEXT, Japan (No. JPMXP0112101001) and JSPS KAKENHI (Nos. 19H05790, 20H00354 and 21H05233).

## Figure captions

Figure 1. (a) Schematic fabrication flow for BN-encapsulated 2D FETs. (b-e) Optical images for the FET fabrication at different steps. In (d) the TMC channel and bottom BN are outlined by white and black dashed lines, respectively. (f) AFM surface morphology for the area indicated by white dashed line in (e). (g) Profile lines indicated in (f). (h) Raman and (i) PL spectra for a typical BN-encapsulated bilayer $WSe_2$ channel before encapsulation and after storage in ambient for one month. Scaler bar: 10 µm.

Figure 2. (a) Forward and backward scanning tests on the transfer characteristics for a typical encapsulated 6L $MoS_2$ FET at 7.5 and 300 K. Insets: enlarged plots to discern the near-zero hystereses. (b) Comparison of the density of trap states in different dielectric environments. Inset: Dependence of hysteresis on $T$ in our $MoS_2$ FET. (c) Arrhenius plots of $I_{ds}/T^{3/2}$ versus $1000/T$ at different gating conditions. (d) Plot of $\phi_B$ versus $V_{TG}$, implying a low barrier height around 34 meV at the flat-band condition. Inset: $R_c$ versus $T$.

Figure 3. (a) Logarithmic and (b) linear plots of transfer curves for an encapsulated 6L $MoS_2$ FET at different $T$s. Inset in (a): optical image for the measured device. (c) Extracted cryogenic $\mu$ versus $T$ with the highest value up to 4272 cm$^2$V$^{-1}$s$^{-1}$ 7.5 K. The solid red line shows the trend of theoretical $\mu$ from CIs by assuming $n_{CI} = 10^{11}$ cm$^{-2}$. (d) Linear and (e) logarithmic plots of transfer curves for an encapsulated 2L $WSe_2$ FET at different $T$s. (f) Extracted cryogenic $\mu$ (electron: red, hole: blue) versus $T$. The red and blue lines show the theoretical $\mu$ trends from CI contribution by assuming $n_{CI} = 1.1 \times 10^{11}$ cm$^{-2}$.

Figure 4. Comparison of (a) $\gamma$ and (b) low-$T$ mobility ($\mu_{LT}$) for $MoS_2$ FETs with different encapsulated dielectrics. Comparison of (c) $\gamma$ and (d) $\mu_{LT}$ for $WSe_2$ FETs with different encapsulated dielectrics. Extracted parameters for hole and electron are shown in blue and red colors, respectively.

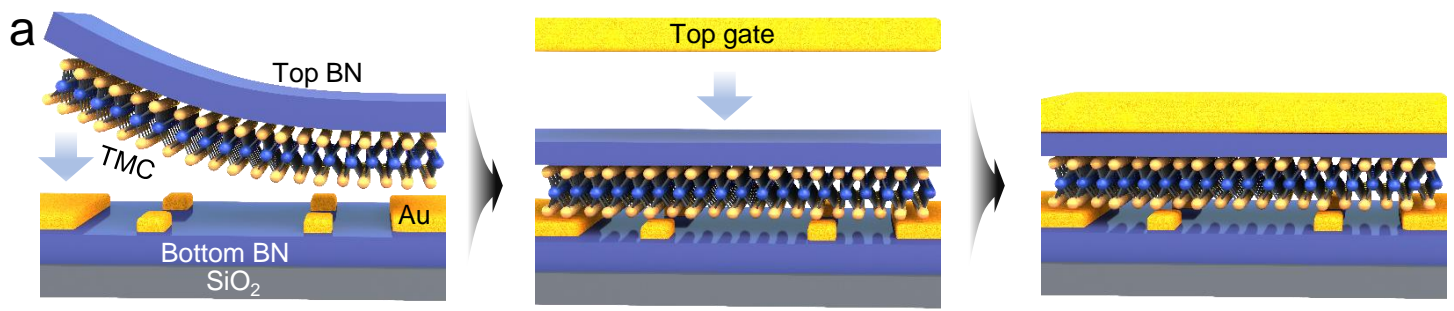

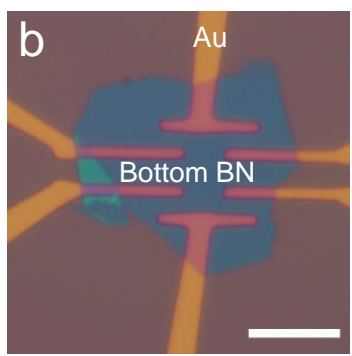 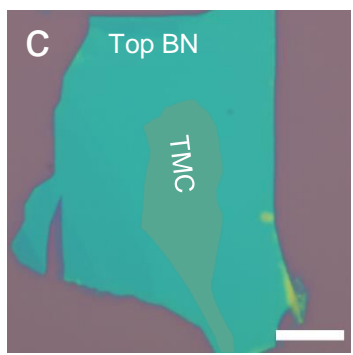 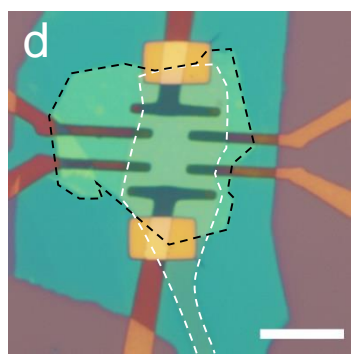 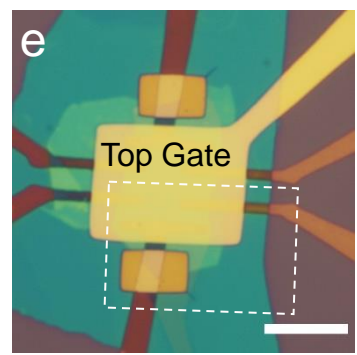

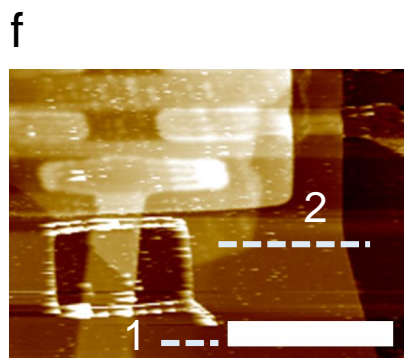 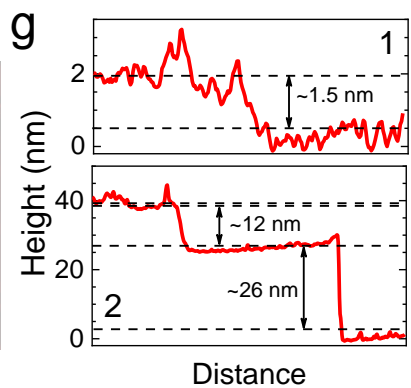 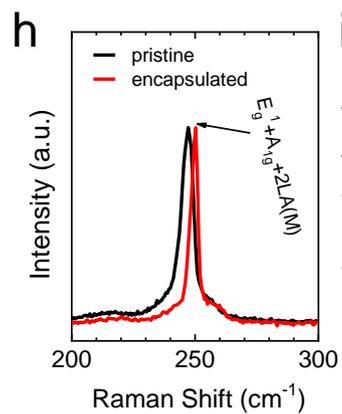 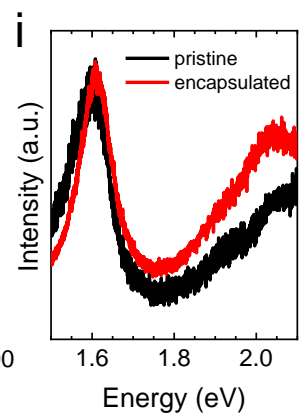

Figure 1

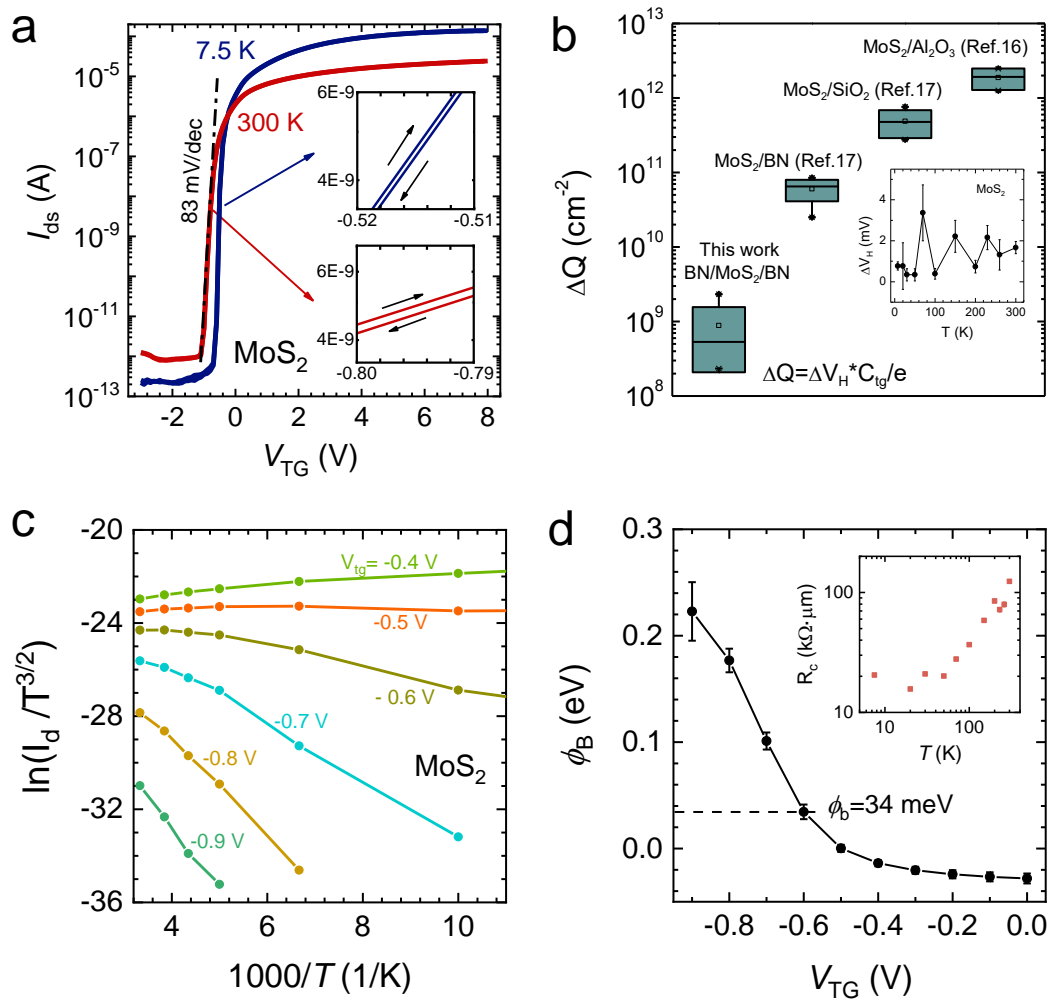

Figure 2

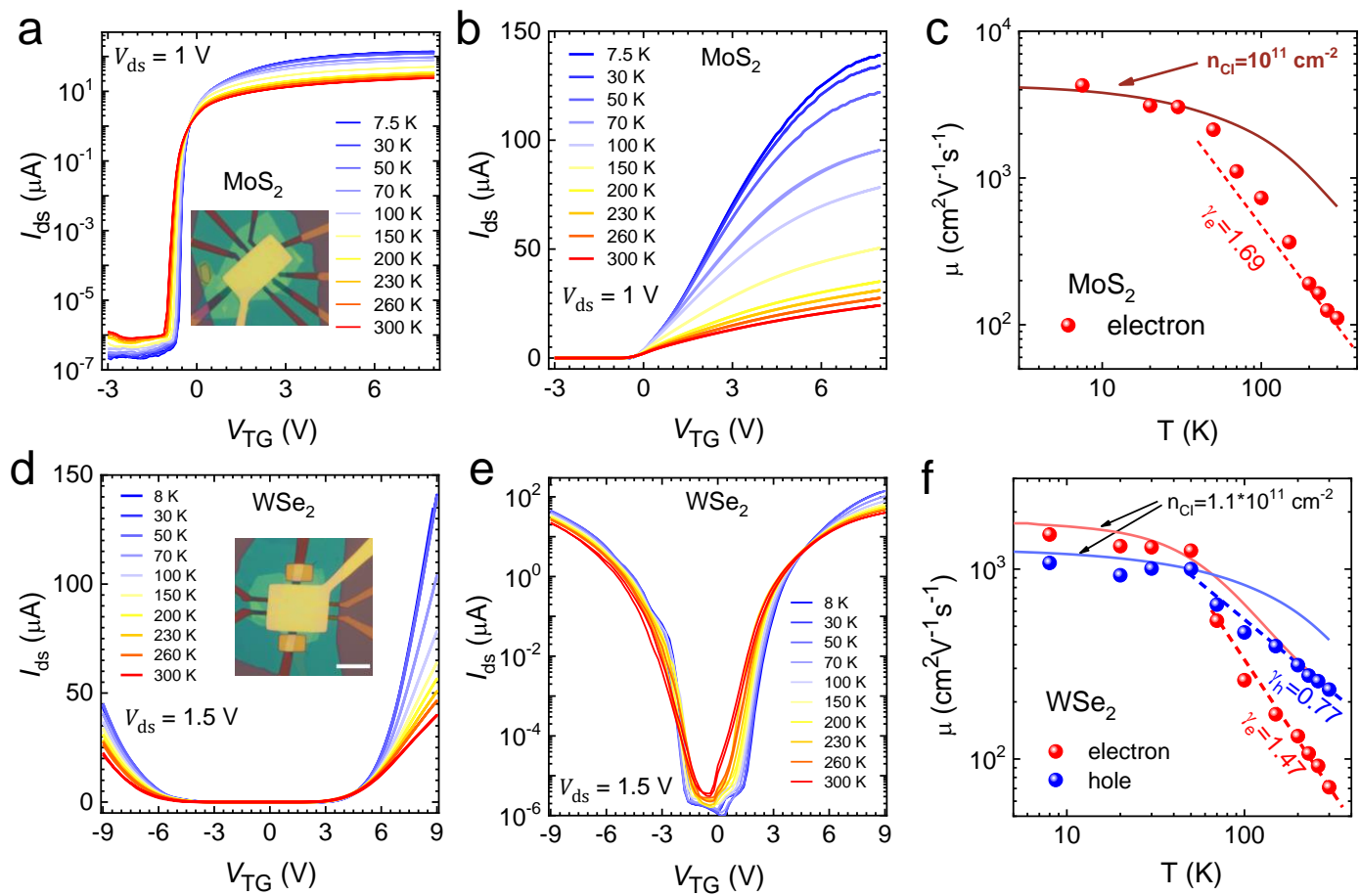

Figure 3

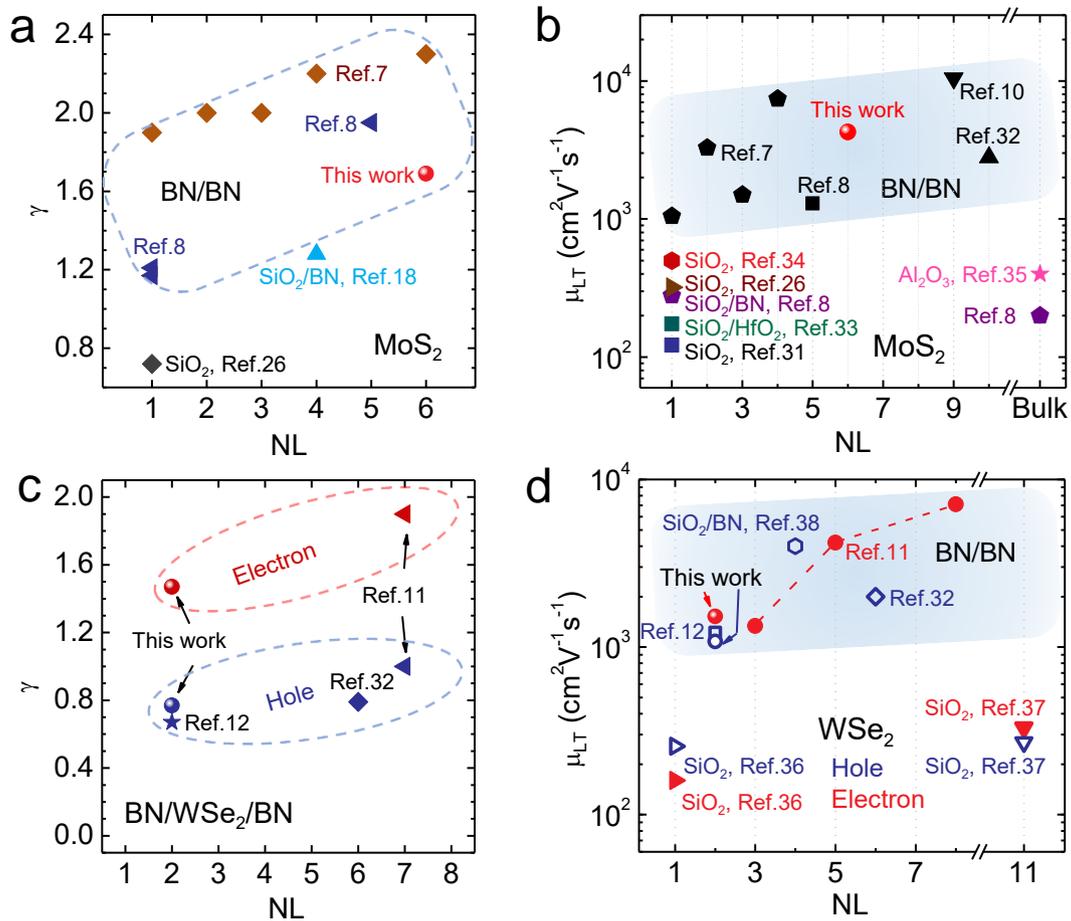

Figure 4